\documentclass[prd,onecolumn,nofootinbib,showpacs,showkeys]{revtex4}
\usepackage{latexsym,graphicx,amssymb,amsmath,mathrsfs}
\usepackage{setspace,bm} 
\newcommand{\be}{\begin{equation}}      
\newcommand{\ee}{\end{equation}}      
\newcommand{\bea}{\begin{eqnarray}}      
\newcommand{\eea}{\end{eqnarray}}    
     
\newcommand{\rt}[1]{{}}

\makeatletter
\renewcommand\appendix{\par
\setcounter{section}{0}%
\setcounter{subsection}{0}%
\gdef\thesection{\appendixname\space\@Alph\c@section}}

\long\def\unmarkedfootnote#1{{\long\def\@makefntext##1{##1}\footnotetext{#1}}}
\makeatother

\begin{document} 

\title{Finite temperature symmetry restoration\\ in the $U_L(3)\times U_R(3)$ linear sigma model\\ from a large-$n$ approximation} 

\author{G. Fej\H{o}s}
\email{fejos@nt.phys.s.u-tokyo.ac.jp}
\affiliation{Department of Physics, The University of Tokyo, Tokyo 113-0031, Japan}
\author{A. Patk{\'o}s}
\email{patkos@galaxy.elte.hu}
\affiliation{Department of Atomic Physics, E{\"o}tv{\"o}s University,
H-1117 Budapest, Hungary}
\affiliation{Research Group for Statistical and Biological Physics 
of the Hungarian Academy of Sciences, H-1117 Budapest, Hungary}

\begin{abstract}
{A recently proposed approximate large-$n$ ground state solution of the $U_L(n)\times U_R(n)$ symmetric linear sigma model is investigated at finite temperature. We study the coupled evaporation of two condensates corresponding to the symmetry breaking pattern $U_L(3)\times U_R(3)\rightarrow U_V(2)$, realized by the ground state in certain parts of the coupling space. The region of the fluctuation induced first order transitions and its second order boundary is mapped out. The existence of a tricritical point is conjectured.}
\end{abstract}

\pacs{12.38.Cy, 11.10.Wx}
\keywords{Chiral symmetry breaking; Large-$n$ approximation}  
\maketitle

\section{Introduction}
The Lagrangian of QCD displays an approximate chiral $U_L(3)\times U_R(3)$ symmetry. Although the axial anomaly reduces the approximate symmetry to
$SU_A(3)\times SU_V(3)\times U_V(1)$, at finite temperature a fast restoration of the $U_A(1)$ factor might give phenomenological relevance to investigations of the $U_L(3)\times U_R(3)$ symmetric linear sigma model. The interest for investigating the symmetry breaking phase structure of the $U_L(n)\times U_R(n)$ symmetric model at large-$n$, and the symmetry restoration pattern at finite temperature stems from the fact that according to the leading order renormalization group analysis, all models with $n\geq 2$ go through a first order symmetry restoration of similar characterization \cite{paterson81,pisarski84}. The fluctuation induced first order nature \cite{coleman73,halperin74} of the transition was confirmed with nonperturbative studies \cite{berges02,berges97a,berges97b,fukushima11,lenaghan00}, while purely perturbative methods \cite{chan73,rosenzweig80}
or large-$n$ approximations based on a $O(2n^2)$ symmetric saddle point \cite{meyer96} could enforce a region of first order transitions only by introducing the explicitly cubic $U_A(1)$ anomaly term of 't Hooft.

It is puzzling that even the most recent lattice simulations failed to establish the region of the explicit symmetry breaking fields where the QCD phase transition displays a first order discontinuity \cite{karsch04,endrodi07,ding11}. Predictions of the effective models concerning the size of this region  fall in the ballpark of the lattice estimations \cite{lenaghan01,herpay05}. In particular, meson models solved in the limit of large flavor numbers represent a promising approach for strongly coupled effective models, when their solution is continued to $n=2,3$ \cite{meyer96,marko10}.

A fundamental obstacle in exploring the nature of the transition with the large-$n$ technique was the absence of any result for the leading order ground state, qualitatively different from the large-$n$ behavior of the $O(2n^2)$ symmetric linear sigma model \cite{jackiw74}.
Recently we have constructed an approximate large-$n$ ground state for the $U_L(n)\times U_R(n)$ symmetric model \cite{fejos10,fejos11} based on the extra assumption that the heavy scalar fields can be treated at low temperatures as static.
Therefore, in the computation of the effective potential, only pseudoscalars were included. This represents a consistent approximation up to temperatures where the fluctuations of the heavy scalars are excited. This feature is increasingly difficult to maintain near the $h_0=0$ axis, where the mass gap between the scalars and pseudoscalars tends to zero at criticality. For this reason the findings of the present report concerning this region should be checked with more complete studies in the future.	

The proposed solution goes beyond the usual 1-loop effective potential, because it is built on a nontrivial saddle point searched in the space of a rather general set of quadratic auxiliary fields: a Lorentz-scalar $U_V(n)$ singlet ($x$), a scalar $U_V(n)$ vector ($y_1^a$), and a pseudoscalar $U_V(n)$ vector ($y_2^a$), introduced in \cite{fejos11} via Hubbard-Stratonovich transformations. Including the propagation of these composite fields, one effectively takes into account diagrams with more than 1 loop in terms of the original field variables, even when considering the 1-loop effective potential in terms of the extended formulation \cite{fejos10}. The symmetry breaking pattern $U_L(n)\times U_R(n)\rightarrow U_V(n-1)$ investigated below is parametrized by two diagonal elements of the $U(n)$ Lie algebra: $v$$=$$\sqrt{2n^2}(v_0{\bf 1}+v_8T_{diag}^n)$, where $T_{diag}^n=diag(1,1,...,1,-n+1)/\sqrt{n(n-1)}$ is the ``longest'' element of the Cartan subalgebra. The choice of this condensate is motivated phenomenologically, since it can be connected the most economically with the $n=3$ case. Below, we consistently use the notation $T_{diag}^n=T^8$ and a similar notation for the corresponding $U(n)$-vector components. The condensate induces nonzero saddle point values for the composite fields $x,y_1^0,y_1^8$. In the absence of any explicit symmetry breaking conjugate to $v_8\equiv\chi_8/\sqrt{n}$, we have found in \cite{fejos11} that the configuration $v_0\neq 0$, $\chi_8=0$ represents the absolute minimum of the effective potential for all values of the quartic couplings $g_1,g_2$ and the external field $h_0$ conjugate to $v_0$, in agreement with common expectations. 

The new observation made in \cite{fejos11} was the existence of a metastable local minimum even in the chiral limit, characterized by $\chi_8\neq 0$, $v_0\neq 0$. The corresponding $v_0$ was the same as in the ground state with the $U_V(n)$ symmetry.
The present investigation focuses on a part of the parameter space, where the explicit breaking $h_8T^8$ drives the newfound local minimum to be the true ground state at low temperatures. We determine the region of the first order transitions and its second order boundary. Characteristic features will be illustrated in a two-dimensional slice, where $h_8$ and the $O(2n^2)$-symmetric quartic coupling are fixed. Qualitative exploration is also outlined varying these two parameters. We speculate that at the edge of applicability of our approximation the observed features imply the existence of a tricritical point. 

The mixing of the composite fields $y_2^a$ with the elementary pseudoscalars results in a mass splitting within this sector. We found $2(n-1)$ mixing heavier pseudoscalars \cite{fejos11}, they are called below ``kaons,'' since they coincide with them at $n=3$. The $(n-1)^2$ light (Goldstone) fields bear the short name ``pion.'' There is an additional heavy $U(n-1)$ singlet analogue of $\eta_{88}$. The $1$-loop effective potential with ${\cal O}(n)$ accuracy, depending on 2 condensates and 3 saddle-point coordinates (i.e. $v_0,\chi_8, x, y_1^0, y_1^8$) has the following expression, when only the pseudoscalar contribution is included (see details in \cite{fejos11}):
\bea
V&=&V_{cl}+V_q,\nonumber\\
V_{cl}&=&n^2\left[M^2v_0^2+\frac{1}{2}(x^2+(y_1^0)^2)-2h_0v_0\right]\nonumber\\
&+&n\left[M^2\chi^2_8+\frac{1}{2}(y_1^8)^2-2h_8\chi_8-i\sqrt{2g_2}y_1^8\chi_8(2v_0-\chi_8)\right],
\nonumber\\
V_q&=&-\frac{i}{2}\left[(n^2-2n)\int_p\ln(-p^2+M_\pi^2)+2n\int_p\ln(-p^2+M_K^2)\right].
\label{auxi-pot}
\eea
One recognizes that the expression $V_{cl}$ contains beyond the leading terms $(\sim n^2)$ also subleading contributions ($\sim n$), which reflect the scaling features of the fields $\chi_8,y_1^8,h_8$. The same distinction in $V_q$ reflects the multiplicities in the pseudoscalar sector. The saddle point equations which determine the values of the auxiliary fields are the following:
\bea
-i\sqrt{\frac{2}{g_1}}x&=&2v_0^2+T^F(M_\pi^2)+\frac{2}{n}\left(\chi_8^2+T^F(M_K^2)-T^F(M_\pi^2)\right)=
 -i\sqrt{\frac{2}{g_2}}y_1^0
,\nonumber\\
-i\sqrt{\frac{1}{2g_2}}y_1^8&=&2v_0\chi_8-\chi_8^2-\frac{1}{2}\big(T^F(M_K^2)-T^F(M_\pi^2)\big),
\label{saddle-eq}
\eea
where $T^F(\mu^2)$ is the finite part of a tadpole integral \cite{fejos10,fejos11}:
\bea
T^F(\mu^2)&=&\frac{1}{16\pi^2}\Big(\mu^2\log\Big(\frac{\mu^2}{\mu_0^2}\Big)-\mu^2+\mu_0^2\Big)+
\frac{T^2}{2\pi^2}\int_{\mu/T}^{\infty}dx \sqrt{x^2-\frac{\mu^2}{T^2}}\frac{1}{e^x-1}.
\label{tadpole}
\eea
The first term of the right-hand side is the renormalized vacuum piece at renormalization scale $\mu_0$ (fixed to $m$ in this study), while the second one is the finite temperature part at temperature $T$. We do not show any counterterms explicitly; the complete renormalization program of the approximation was presented in \cite{fejos11}. The masses $M_\pi^2$ and $M_K^2$ appearing in (\ref{auxi-pot}) and (\ref{saddle-eq}) are determined by the gap equations
\be
M_\pi^2=M^2 -\frac{i}{n}\sqrt{2g_2}y_1^8,\qquad M_K^2=M^2+i\sqrt{\frac{g_2}{2}}y_1^8+g_2\chi_8^2,
\ee
where
\bea
M^2&\equiv& -m^2-i(\sqrt{2g_1}x+\sqrt{2g_2}y_1^0)=-m^2+(g_1+g_2)\left[2v_0^2+T^F(M_\pi^2)+\frac{2}{n}\left(\chi_8^2+T^F(M_K^2)-T^F(M_\pi^2)\right)\right].
\label{gap-eq}
\eea
 The ${\cal O}(n^0)$ part of (\ref{gap-eq}) coincides with the gap equation of the $O(2n^2)$ symmetric model up to the missing scalar contribution, omitted by the heaviness assumption. Only the ${\cal O}(1/n)$ contributions introduce intrinsic $U_L\times U_R$ breaking features. (Note the change of the $m^2$ sign convention relative to \cite{fejos11}). The strength of all these terms is governed by $g_2$ alone, which appears in the masses figuring in the ${\cal O}(1/n)$ piece of this equation. This is made explicit also in (\ref{v-EoS}) below.  
Exploiting the saddle point equation of $y_1^8$ one obtains a gap equation for $M_K^2$ which determines directly its dependence on $v_0$ and $\chi_8$:
\be
M_K^2=M^2-2g_2\left(\chi_8(v_0-\chi_8)-\frac{1}{4}\big(T^F(M_K^2)-T^F(M_\pi^2)\big)\right).
\label{mk-gap}
\ee

\section{Application of the large-$N$ potential at finite $N$}

The $v_0\neq 0, \chi_8\neq 0$ nontrivial solution originally was found in a strict large-$n$ study by separately minimizing the ${\cal O}(n^2)$ and the ${\cal O}(n)$ parts of the potential (\ref{auxi-pot}) \cite{fejos11}. Then the leading order equation describes the restoration of the $O(2n^2)\subset U_L(n)\times U_R(n)$ symmetry. In the subleading equations, one searched for minima with respect to $\chi_8$ in the background of $v_0$ determined from the leading ${\cal O}(n^2)$ potential. In the present paper we shall consider the case $n=3$, and search for stationary points of the sum $V$ (both at $T=0$ and at nonzero temperatures). 

For the evaluation of the sum we use the ${\cal O}(1/n)$ accurate values of $v_0,M^2,\chi_8,M_K^2$. Since in the equations of $\chi_8$ and $M_K^2$ only the ${\cal O}(n)$ part of the potential is involved, those equations are unchanged as compared to the equations used in our previous paper \cite{fejos11}. Namely, in addition to (\ref{mk-gap}), one has the following equation of state for $\chi_8$ after substituting $y_1^8$ from (\ref{saddle-eq}) into the derivative of $V$ with respect to $\chi_8$:
\be
0=h_8-M^2\chi_8-2g_2(v_0-\chi_8)\left[\chi_8(2v_0-\chi_8)-\frac{1}{2}\big(T^F(M_K^2)-T^F(M_\pi^2)\big)\right]-g_2\chi_8T^F(M_K^2).
\label{chi-EoS}
\ee
The equation of state of $v_0$, including also the ${\cal O}(1/n)$ contributions, looks like
\be
0=h_0-M^2v_0-\frac{2g_2}{n}\chi_8\left(2v_0\chi_8-\chi_8^2-\frac{1}{2}\left(T^F(M_K^2)-T^F(M_\pi^2)\right)\right).
\label{v-EoS}
\ee

For the realization of the strategy outlined above, Eqs. (\ref{gap-eq}) and (\ref{v-EoS}) were solved in the spirit of the large-$n$ approximation iteratively, by substituting into them
\be
M^2=M_0^2+\frac{1}{n}M_1^2,\qquad v_0=v_{00}+\frac{1}{n}v_{01}.
\label{n-expansion}
\ee
In (\ref{mk-gap}) and (\ref{chi-EoS}) it is sufficient to replace everywhere $M_\pi^2$ (and $M^2$) by $M_0^2$ and use $v_{00}$. These quantities (together with $M_1^2$ and $v_{01}$) are found from the order by order solution
of Eqs. (\ref{gap-eq}) and (\ref{v-EoS}) keeping the first two orders of their respective power series in $1/n$. The values of $v_0$ and $M^2$ are obtained then from (\ref{n-expansion}), where we put $n=3$. These values are used together with $\chi_8$ and $M_K^2$ when one searches for the minima of the potential (\ref{auxi-pot}) as a function of the temperature.

\section{Phase diagram from the numerical solution}

Our interest in the numerical investigation was to see if the two condensates $v_0$ and $v_8\equiv \chi_8/\sqrt{3}$ evaporate strongly correlated enough to expect a single unique transition in a certain part of the coupling space. We consider only the case when at $T=0$ the "nontrivial" $\chi_8\neq 0$ minimum, not proportional to $h_8$ represents the true ground state. 
The complete exploration of the four-dimensional parameter space ($h_0,h_8,g_1,g_2$) is beyond the present scope. Since $v_{00}$ depends only on $g_1+g_2$, we decided to fix this sum. The leading order solutions $M_{00}^2,v_{00}$ were therefore tuned exclusively by $h_0$. We varied $g_2$ at $T=0$  in a large interval, where the approximate symmetry of the ground state is broken down to $U_V(2)$, if a large enough value is fixed for $h_8$, and $h_0$ is near zero. With the choice of $h_8=0.04m^3$ and $g_1+g_2=14.5$, we found exclusively positive squared masses in the whole $v_8$ interval of interest, avoiding any problem related to a complex effective potential \cite{rivers87,weinberg87}. We shall present the results for this case in some detail, followed by statements concerning the effect of the variation of $h_8$ and $g_1+g_2$.

Two further observations set upper bounds on the variation of $g_2$ and $h_0$. For $g_2>g_{2max}=14.5$, the system becomes unstable, since in the stability region $g_1>0$. Furthermore, when one increases $h_0$ at fixed $g_2$, one encounters a value where the strength of $h_8$ is not sufficient anymore to drive the nontrivial solution of $v_8$ to be the absolute minimum at $T=0$. 
\begin{figure}
\centerline{\includegraphics[bb=50 50 554 770,width=7.1cm,height=11.7cm,angle=270]{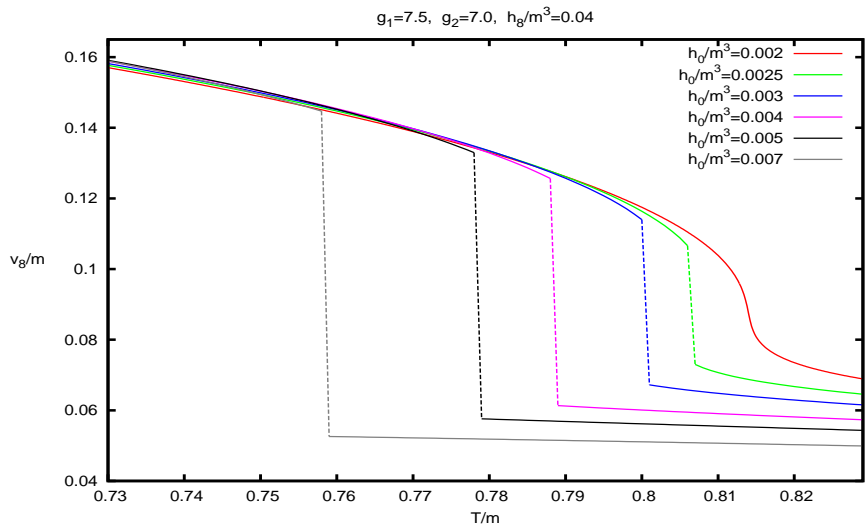}}
\caption{The critical value of $h_{0,c}$ is found as the vanishing point of the discontinuity in $v_8$ when $h_0$ is varied. The curves from left to right display the discontinuities belonging to decreasing $h_0$ values. The critical value is fitted to $h_{0,c}\approx 0.00211m^3$.}
\label{Fig:1}
\end{figure}

The region of first order transitions restoring the approximate $U_V(3)$ symmetry was mapped out in the ($g_2,h_0$) plane by observing a discontinuity of $v_8$ at some transition temperature and nonvanishing value of $v_0$. 
The size of the discontinuity $\Delta v_8$ increases with $g_2$ at fixed $h_0$. This tendency is the same as found with exact renormalization group studies in the $n=2$ case \cite{berges02,berges97a,berges97b,fukushima11}. In the opposite direction, one finds a critical value $g_{2,c}(h_0)$ below which the transition becomes a crossover. 

\begin{figure}
\centerline{\includegraphics[bb=50 50 554 770,width=7.1cm,height=11.4cm,angle=270]{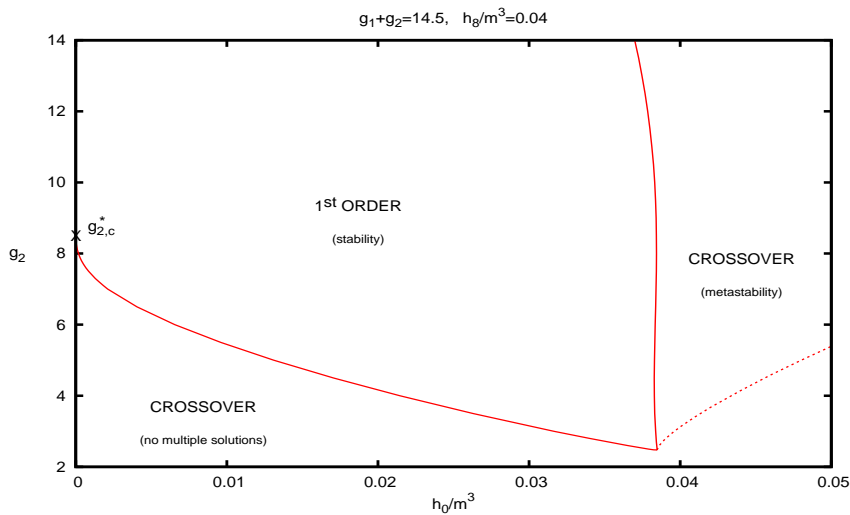}}
\caption{Boundary lines of the first order transition region in the $g_2-h_0$ plane. The quasihorizontal curve shows the second order boundary of the first order transition, while the quasivertical one delimits the region where $h_8=0.04m^3$ is sufficient to push the nontrivial local minimum into the global minimum of the effective potential. The dashed line only signals the boundary of the existence of the nontrivial ($v_8\neq 0$) local minimum at $T=0$.}
\label{Fig:2}
\end{figure}

The edge of the first order transition region was localized by looking for the point where $\Delta v_8$ vanishes. $h_{0,c}(g_2)$ was found at fixed $g_2$ ($2.5 \lesssim g_2 \lesssim 8.5$), by varying $h_0$. Figure \ref{Fig:1} illustrates the method. In another approach, we used that continuous transitions are signaled by the $\lambda$ singularity of an appropriate susceptibility. Remarkably, $v_0$ reacts sensitively to the restoration of the approximate $U_V(3)$ symmetry becoming continuous. Namely, $dv_0/dh_0$ displays a well-expressed $\lambda$ singularity, although this quantity is not generically related to the $U_V(2)\leftrightarrow U_V(3)$ change in the approximate symmetry pattern. This behavior reflects a rather strong coupling between the two order parameters. The first method locates $h_{0,c}(g_2)$ with an accuracy ${\cal O}(10^{-5})$, while determination of the same curve as $g_{2,c}(h_0)$ was consistent with it. In Fig. \ref{Fig:2} we display the points of the boundary of the first order transition region. The boundary which separates regions with three resp. two minima of the effective potential at $T=0$ is drawn by the dashed line. Correspondingly, it is possible to have the nontrivial $\chi_8\neq 0$ minimum as a stable or metastable state. In the latter case the phase evolution proceeds from the ``trivial'' $v_0\neq 0, \chi_8\sim h_8$ minimum into the ``fully trivial'' minimum where $v_0, v_8$ are both proportional to the explicit breaking fields.

A high quality fit to the phase boundary of the scaling form 
\be
h_{0,c}=a(g_2-b)^y,\qquad a=(8.70\pm0.25)\cdot 10^{-4},\qquad b=8.49\pm0.01,\qquad y=2.23\pm0.02
\label{Eq:phase-boundary}
\ee
shows that the phase boundary crosses the $h_0=0$ axis at $g^*_{2,c}\approx 8.5$. The fit is mostly sensitive to the boundary points near the $h_0=0$ axis, therefore, in its neighborhood the transition points were determined quite densely. The errors vary with the chosen points whereupon the fit was made. To obtain (\ref{Eq:phase-boundary}), we chose the region $g_2\in [6.5,8.4]$. Errors of the exponent should be estimated from the variation of the fitted parameters with respect to the chosen region of the points in the first place, not just by root-mean-square errors of a single fit. It is quite remarkable that the value of the critical exponent $y$ is very close to the value $5/2$, which reminds us the reciprocal of the mean field exponent characterizing the quark-mass/external field dependence of various quantities (e.g. chemical potential, critical temperature, or couplings) near a tricritical point \cite{hatta03,herpay06}. In the suspected tricritical point then
one would expect that a second order line along the $h_0=0$ axis starting from the origin will meet the one displayed in Fig. \ref{Fig:2}, which would indicate continuous transition of the two nontrivial minima into the fully trivial symmetric state at the tricritical point.
Moreover, there is an additional interesting indication pointing to an increasingly correlated evaporation of $v_0$ and $v_8$ near $h_0=0$. 

\begin{figure}
\centerline{\includegraphics[bb=50 50 554 770,width=6.6cm,height=11.4cm,angle=270]{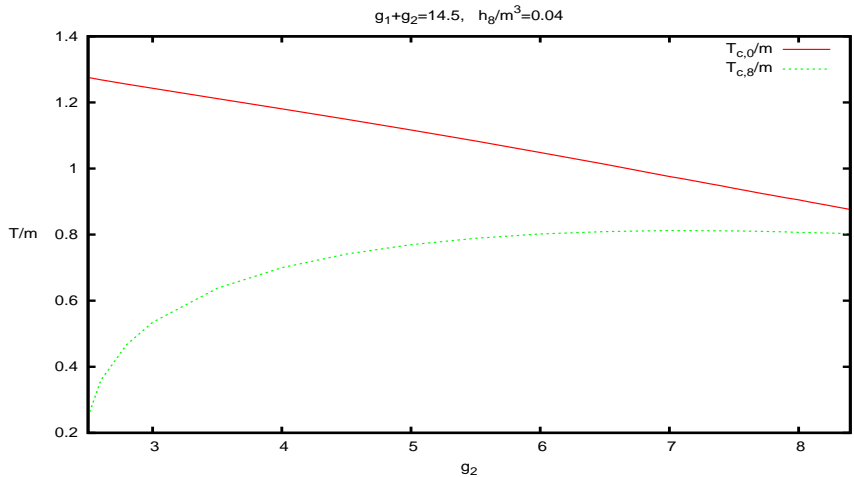}}
\caption{The variation of the transition temperature of $v_8$ along the phase boundary parametrized by $g_2$. The lower curve comes from the zero of the discontinuity $\Delta v_8$  and the lupper from the maximum of $dv_{00}/dh_0$ (which depends on $g_2$ only due to the $g_2$-dependence of the phase boundary line).}
\label{Fig:3}
\end{figure}

Figure \ref{Fig:3} shows the variation of the critical temperature $T_{c,8}(g_2)$ along the phase boundary and the pseudocritical temperature $T_{c,0}(g_2)$. The latter characterizes the $v_0$ evaporation from the ``trivial'' into the ``fully trivial'' minimum, determined by the maximum of $dv_{00}/dh_0$ evaluated at $h_{0,c}(g_2)$. The two curves approach each other very close near the $h_0$ axis. It shows that the transition of $v_8$ from the nontrivial into the trivial ($v_8\sim h_8$) minimum precedes the transition point of $v_{00}$, and the $v_8$ jump is followed by a separate crossover in $v_{00}$ accompanied by the further gradual decrease of $v_8$. This corresponds to the sequential restoration $U_V(2)\rightarrow U_V(3)\rightarrow U_L(3)\times U_R(3)$.

With the increase of $h_8$ we observe an increase in $g^*_{2,c}$ and the two curves on Fig. \ref{Fig:3} further approach each other. This tendency continues monotonically until $g^*_{2,c}=14.5$ is reached, but the difference of the critical temperatures ($\Delta T$) does not change sign. Further increasing $h_8$ pulls away the critical point from the $h_0$ axis along the $g_2=14.5$ line with increasing $\Delta T$. The analysis can be repeated with increasing $g_1+g_2$ values and along the $h_0$ axis. One finds diminishing minimum values for $\Delta T$ reached always at $g_1=0, g_2=g_{2,c}$. The corresponding $h_8$ gets smaller as we increase $g_1+g_2$. It is at $g_1+g_2\approx 30.8$ where the minimum of $\Delta T$ becomes zero (we have $h_8\approx 0.037m^3$ at this point). Going higher in $g_1+g_2$ one even finds that the restoration of $v_0$ precedes the evaporation of $v_8$. This clearly shows that in this region the one-step recovery of the approximate full symmetry from approximate $U_V(2)$ does occur.

The last issue to be discussed is the validity of the assumed mass-hierarchy in the temperature range of the investigation. The range of temperatures involved in the study does not exceed the mass scale $m$. For the squared mass-ratios in the interior of the first order region, we found that the temperature dependent behavior of the pion and kaon mass is compatible with neglecting the heavy scalars. However, in points near the boundary of the first order region and close to $h_0=0$ the scalar-pseudoscalar mass difference parametrically proportional to $v_0^2$ decreases when the temperature is close to the critical value. Therefore just in the region of the conjectured tricritical behavior our results should be checked with an improved treatment including the scalar fluctuations, too.

In summary, for fixed values of $h_8$ and $g_1+g_2$ we determined the boundary between the region of the crossover and the first order transition restoring the approximate $U_V(3)$ symmetry and found evidence for a strongly coupled behavior of the condensates $v_0$ and $v_8$ near the $h_0\rightarrow 0$ $(h_8\neq 0)$ limit. Tendencies observed when $h_8$ and $g_1+g_2$ are varied appropriately led us to a region of single step $U_V(2)\rightarrow U_L(3)\times U_R(3)$ transition.
   
\section*{Acknowledgements}
The authors are grateful to A. Jakov\'ac and Zs. Sz\'ep for their careful reading of the manuscript and valuable suggestions. This work is supported by the Hungarian Research Fund under Contracts No. T068108 and K77534. G. F. is supported by the Japan Society for the Promotion of Science under ID No. P11795.

\end{document}